\documentclass[singlespacing]{elsart}
\usepackage{graphics}
\usepackage{amssymb}
\begin{document}
\begin{frontmatter}
\title{Study of the one dimensional Holstein model using the augmented space approach}
\author[Atis]{Atisdipankar Chakrabarti}
\address[Atis]{R.K.M.V.C. College, Rahara, 24 Parganas (North), West Bengal, India}
\author[Bose]{Monodeep Chakraborty}
 \author[Bose]{A. Mookerjee}
\address[Bose]{S.N. Bose National Centre for Basic Sciences, JD Block, Sector III, Salt Lake, Kolkata 700098, India}
\begin{abstract}
A new formalism using the ideas of the augmented space recursion (introduced
by one of us) has been proposed to  study  the ground state properties of ordered and disordered
one-dimensional Holstein model. For ordered case our method works equally 
well in all parametric regime
and matches with the existing exact diagonalization and DMRG results. On the other hand
the quenched substitutionally disordered model works in low and intermediate regime of electron phonon coupling.
Effect of phononic and substitutional disorder are treated on equal footing.

\end{abstract}
\end{frontmatter}

\section{ Introduction}
When a conduction electron or hole moves through a polar crystal it polarizes and
distorts the neighboring ion lattice sites. When it moves 
it carries along with it the lattice distortion. The electron together with this lattice
distortion or self-consistent polarization field is a quasi particle and is referred
to as polaron. Many experiments indicate strong evidence of polaronic carriers in strongly
correlated electronic materials like manganites (showing colossal magneto-resistance) 
\cite{cmr1}-\cite{cmr2}, organic \cite{org1} and high $T_c$ superconducting
cuprates \cite{sc1}-\cite{sc2}. Holstein model\cite{hoo} has been one of the
basic model used to study the formation and nature of polaron.
Analytical approaches to solve the model have got limited success.
Main draw back of those results are they are valid only in a 
limited range of parametric regime of electron-phonon coupling \cite{mott1}
and also for small system size. 

Recently there has been a spurt in numerical studies on this model, which
includes the variational approach  based on exact 
diagonalization method (VAED) \cite{trug1}-\cite{trug2}, the 
density-matrix renormalization group
 techniques (DMRG) \cite{dmrg1}, exact diagonalization techniques(ED) \cite{ed1}-\cite{ed7}, the quantum Monte
 Carlo calculations (QMC) \cite{qmc1}-\cite{qmc2} and global-local method \cite{glm1}. None of these methods work equally well in all parametric regimes. Each has some shortcomings. VAED, the most recent and successful of the
 methods, 
though highly accurate in the the weak and intermediate coupling regimes, 
 fails to maintain its accuracy in the strong coupling limit. The
DMRG method is  accurate in the strong and intermediate coupling regimes,
but is not accurate in the weak coupling regime. The DMRG results are
 not translationally symmetric. 
	DMRG lacks the accuracy of VAED in the weak coupling regime because
its lattice size is smaller than the spatial extent of the large polaron that will 
be formed. The VAED cannot emulate its success of the low and intermediate 
regime in the the strong coupling regime because the electron-phonon system 
demands more phonons in its `root state' \cite{trug2}
 and its near vicinity. 

 Disordered Holstein model has been earlier attempted by \cite{bronol}
 using Dynamical CPA. But their method is effective in the intermediate 
electron-phonon coupling regime and for only infinite coordination, i.e. 
on a Bethe Lattice. 
Keeping this in mind, in this communication 
we have proposed a reciprocal space recursion technique, introduced earlier by us in a different context \cite{am3}, as a feasible method for the study of the Holstein model. We have constructed a basis which meets the requirement 
in {\sl all} parametric regimes. 

\section{The Augmented Space Formalism for ordered Holstein model}

In this section we shall describe the setting up of the electron-phonon Hamiltonian as an operator
on the underlying Hilbert space. The electronic part will be represented in a tight-binding like
basis. 
In our simple model there  will be 
only a single orbital per site and the tight-binding basis will be  labeled by a site $R_n$ : $\{\vert R_n\rangle\}$. This basis
will span the electronic part of the Hilbert space ${\bf H}^{el}$. The phonon
part will be described by a {\sl configuration state} which will indicate how many phonon excitations there
are at each {\sl site} labeled by $R_n$. Each {\sl configuration} will be a pattern of the type 
$\{ n_{R_1}, n_{R_2} \ldots n_{R_n} \ldots\}$ where $n_{R_n}$ takes the values $0,1,2,\ldots$
and is called the {\sl cardinality} of the site $R_n$. The pattern is called the {\sl cardinality
sequence} and uniquely labels a phonon state. These states will span the phonon part of the Hilbert space
${\bf H}^{ph}$. The total Hilbert space of states of the electron-phonon system will be ${\bf H}\ =\ 
{\bf H}^{el}\otimes{\bf H}^{ph}$. A member of the basis in this product space
will be labeled {\sl both} by the site labeled real-space part and a configuration
part which tells us how many local phononic excitations are there at {\sl each}
site in the lattice, e.g. $\vert R_n\otimes \{n_{R_1},n_{R_2},\ldots\}\rangle$.

In this basis, the Holstein Hamiltonian \cite{hoo} is :

\begin{eqnarray}  H =\ \varepsilon \sum_{R_n} P_{n}\otimes {I} -t \sum_{R_n}\sum_{R_m} T_{nm}\otimes\ { I}
+\hbar\omega \sum_{R_n} I\otimes{ N}_{n}-\lambda
\sum_{R_n}  P_n \otimes \left({ T}_n^+
 + { T}_n^-\right) 
\end{eqnarray}

\noindent where,
\begin{itemize}
\item $P_n$ is the projection 
operator $\vert R_n\rangle\langle R_n\vert$ 
and  $T_{nm} = \vert R_n\rangle\langle R_m\vert$ 
is the transfer operator in the electronic part of the
Hilbert 
space ${\bf H}^{el}$,
\item   ${ T}_{n}^\pm= 
\vert \ldots n_{R_n}\pm 1\ldots\rangle
\langle \ldots  n_{R_n}\ldots\vert$ are the step up or step down operators 
and ${ N}_n $ is the number operator  
$n_{R_n} \vert \ldots n_{R_n}\ldots\rangle\langle \ldots 
n_{R_n}\ldots\vert$ both in the phononic part of the Hilbert
space ${\bf H}^{ph}$.
\item Here a single electron with an on-site energy $\varepsilon$ can hop to its near neighbour on an infinite lattice with hopping integral $t$
and interacts with dispersion less optical phonons with frequency $ \omega $. The
 electron-phonon coupling strength is denoted by  $\lambda$ . The parameters $\varepsilon$,$t$, $\hbar\omega$
and $\lambda$ have the units of energy.
\item  In all subsequent calculations energies are
scaled in units of t.
\end{itemize}

 This type of Hamiltonian in an augmented space ${\bf H}^{el}\otimes{\bf H}^{ph}$ has been introduced earlier by us \cite{mook} to deal with time dependent disorder in an electronic system, precisely of the form produced by lattice vibrations. The  great advantage of this
formulation is that if we have quenched disorder in $\varepsilon(R_n)$, then a simple augmentation of the Hilbert 
space ${\bf H}^{el}$ by the configuration space of the $\varepsilon(R_n)$ allows us to deal with all the
configuration averaged properties of a disordered Holstein model. This extension will be carried out
by us in a later section. 
Since we are interested in calculating the spectral function, dispersion curves of the system, we
generalize the above mentioned formalism to generate basis states in reciprocal space.
Here a general state in the augmented reciprocal-space basis has the following form : 
\[
|k\otimes\{n_{R_1},n_{R_2}\ldots n_{R_n}\ldots\} \rangle = (1/\sqrt{N})\sum_{R_n} \mbox{exp}(ik\cdot R_n)
|R_n\otimes \{n_{R_1},n_{R_2}\ldots n_{R_n}\ldots\}\rangle.\]

Here $k$ is a reciprocal space vector and  we have Fourier transformed the electronic part of the basis to momentum space keeping the bosonic
part unchanged. In the recursion method, which we shall later employ, we shall
 generate the basis states by repeatedly applying the Hamiltonian on the following 
starting state : 

\[\vert 1\rangle = \vert k\otimes \{\emptyset\}\rangle = (1/\sqrt{N})\sum_{R_n} \exp\{i k\cdot R_{n}\}
\vert R_n\otimes \{\emptyset\}\rangle\] 

Here {$\{\emptyset\}$} denotes a zero phonon configuration in the bosonic subspace of the full Hamiltonian. That is, with a null cardinality sequence.

Alternatively, we can generate basis states recursively 
by similar action of the Hamiltonian on the following two 
states :
 
\begin{enumerate}
\item[(i)] \[\vert 1\rangle = \vert k\otimes \{\emptyset\}\rangle = (1/\sqrt{N})\sum_{R_n} \exp\{i k\cdot R_{n}\}
\vert R_n\otimes \{\emptyset\}\rangle\]
\item[(ii)]
\[\vert 2\rangle = \vert k\otimes \{N_{1},0,0\ldots\}\rangle = (1/\sqrt{N})\sum_{R_n} \exp\{i k\cdot R_{n}\}
\vert R_n\otimes \{N_{1},0,0\ldots\}\rangle\]  
\end{enumerate}

Here ${N_{1},0,0..}$ denotes $N_{1}$ number of phonon at site $1$ of the bosonic subspace of the Hamiltonian and zero else where. We shall call this the phonon-enriched state. 

The advantage of the  second choice is that, if we apply the Hamiltonian $N$ times on the two initial
states, $N+N_{1}$ number of phonons are generated in the site $1$ and sufficient
number of bosons are generated in its vicinity, while in the first choice we get only $N$ number
of phonons in the first site and progressively less numbers in the neighboring sites.
This is important in order to get a convergent result in the high e-ph coupling regime.

 In order to calculate averaged spectral function we 
tri-diagonalize the Hamiltonian using the following three
term recursion ~:

We start from a state $\vert k\otimes\{n_{R_1},\ldots\}\rangle\ =\ \vert 1\rangle$ and generate the other states through the following three term recursion~:

\begin{equation}
 \vert n+1\rangle = H\vert n\rangle - \alpha_n(k)\vert n\rangle - \beta ^2_{n-1}(k)\vert n-1\rangle
\end{equation}
                                                                                            
The coefficients $\alpha_n(k)$ and $\beta^2_n(k)$ are obtained by ensuring that the new basis
generated is orthogonal. This yields~:
                                                                                            
\begin{eqnarray*}
 \alpha_n(k) =\langle n\vert H\vert n\rangle /\langle n\vert n\rangle 
 \quad\quad \beta_n^2(k) = \langle n+1\vert n+1\rangle/\langle n\vert n\rangle
\end{eqnarray*}

To carry out the recursion we need to know
the action of the Hamiltonian (1) on a general {\sl state}~:
If write the Hamiltonian (1) as : $H = H_1+H_2+H_3+H_4$ then
\begin{eqnarray}
H_1\vert k\otimes \{n_{R_1},n_{R_2}\ldots \}\rangle & =&
\varepsilon \vert k\otimes \{n_{R_1},n_{R_2}\ldots \}\rangle \nonumber\\
H_2\vert k\otimes \{n_{R_1},n_{R_2}\ldots \}\rangle
& =&
-t\sum_\chi \exp\{ik\cdot \chi\}\vert k\otimes
\{n_{R_1-\chi},n_{R_2-\chi}\ldots \}\rangle\nonumber\\
H_3\vert k\otimes \{n_{R_1},n_{R_2}\ldots \}\rangle & =&
\left(\rule{0mm}{3mm} \hbar\omega \sum_n n_{R_n}\right) \vert
k\otimes \{n_{R_1},n_{R_2}\ldots \}\rangle \nonumber\\
H_4\vert k\otimes \{n_{R_1},n_{R_2}\ldots \}\rangle & =&
-\lambda \vert k \otimes \{n_{R_1}\pm 1, n_{R_2},\ldots\}\rangle
\end{eqnarray}
                                                                                            
Once we carry out a recursive determination of the coefficients $\{\alpha_n(k), \beta^2_n(k)\}$, the Green function is given as a continued fraction~:
                                                                                            
\[ G(k,E) = \frac{1}{\displaystyle E-\alpha_1(k)-
                   \frac{\beta_1^2(k)}{\displaystyle E-\alpha_2(k) -
                   \frac{\beta_2^2(k)}{\displaystyle
{{\ddots}\atop{\displaystyle{\frac{\beta_{N-1}^2}{\displaystyle{E-\alpha_N(k)-\beta_N^2(k)\ T(k,E)}}}}}
}}}
\]
 We carry out recursion up to a finite number of N steps and then terminate the continued fraction with a herglotz terminator $T(k,E)$ \cite{her} suggested by Beer and Pettifor \cite{bp}. The spectral function is then given by :
                                                                                            
\[  A(k,E) = -(1/\pi) \Im m \ll G(k,E)\gg \]

The dispersion curves are obtained by fitting Lorentzian's  in the neighbourhood of
 the peaks in the spectral function : the Lorentzian centre gives the energy
$E(k)$ while the width gives the width in the dispersion curves.

\par We apply the recursion based conjugate gradient (CG) technique \cite{vm}
to obtain  ground state energy and wave function.
To calculate the excited state wave function we  find a orthogonal state
to  the ground state using Gram-Schmidt method of orthogonalization and then
apply CG technique to find the first excited state energy and wave function.
With the wave function and dispersion curve
at our disposal we find the various correlation functions,effective mass from it.
The effective mass we have calculated using the standard formula :

\begin{equation}
\frac{m_{0}}{m_{*}} = \left. \frac{1}{2t} \frac{\partial^{2}E(k)}{\partial k^{2}} \right|_{k=0}
\end{equation}
The mean phonon number indicates a measure of the phononic character of the
polaron.
\begin{equation}
N^{ph} \ =  \langle \psi_{0}|I\otimes{N}_n|\psi_{0}\rangle
\end{equation}

Here $\vert\psi_0\rangle$ is the ground state wave function obtained from the CG technique. 
  The static correlation function between the electron position and oscillator
displacement is given

\begin{equation}
\chi(R_i-R_j)= \langle \psi_{0}|P_i\otimes({T}_{j}^++{T}_{j}^-)|\psi_{0}\rangle
\end{equation}

 The number of excited phonons in the vicinity of the electron is given by

\begin{equation}
\gamma(R_i-R_j)= \langle \psi_{0}|P_i\otimes {N}_j|\psi_{0}\rangle
\end{equation}

\section{Augmented space formalism for disordered Holstein model}
In this section we shall generalize our ideas to a disordered Holstein model.
We shall start with a Hamiltonian :

 \begin{eqnarray}  H =\  \sum_{R_n}\varepsilon(R_n) P_{n}\otimes {I} -t   \sum_{R_n}\sum_{R_m} T_{nm}\otimes\ {I}
  +\hbar\omega \sum_{R_n} I\otimes{N}_{n}\nonumber -\lambda
      \sum_{R_n}  P_n \otimes \left({T}_n^+
      + {T}_n^-\right)
     \end{eqnarray}

\noindent where,
\[ \varepsilon(R_n) = \varepsilon_A\ n(R_n) + \varepsilon_B\ (1-n(R_n))
\]
                                                                                
Here $n(R_n)$ is a random variable which takes the value 0 if $R_n$ is occupied by an A type of  atom and 1 if it is occupied by a B type. The probability of these events are $x$ (the concentration of A atoms in the
alloy) and $1-x$ (concentration of B atoms in the alloy) respectively.
IN this simple model we shall consider only this type of binary 
substitutional disorder. For real disordered alloys the other parameters $t$, $\omega$ and $\lambda$ may also be  random.
The augmented space formalism for
quenched disorder \cite{asf} writes the probability density
of the random variables as :
                                                                                
\begin{eqnarray*}
 { P}[n(R_n)]& =& x\ \delta[n(R_n) -1] + (1-x)\ \delta[n(R_n)] \\
\phantom{X}& & \\
& = & -(1/\pi)\ \Im m\ \langle \uparrow\vert \left((n(R_n)+i\delta){ I}
- { M}^{(n)} \right)^{-1}\vert \uparrow\rangle
\end{eqnarray*}
                                                                                
The operator { M}$^{(n)}$ associated with the random variable $n(R_n)$ is
such that its spectral density is the probability density of the random variable. Here the representation of this operator is
                                                                                
\[ { M}^{(n)} \ =\ \left( \begin{array}{cc}
                                 x & \sqrt{x(1-x)} \\
                               \sqrt{x(1-x)} & 1-x \end{array} \right) \]
                                                                                
\noindent in the basis $\vert\! \uparrow \rangle = \sqrt{x}\vert 0\rangle +
\sqrt{1-x}\vert 1\rangle $  and $\vert\! \downarrow \rangle = \sqrt{1-x}\vert 0\rangle -
\sqrt{x}\vert 1\rangle $, where $\vert 0\rangle$ and $\vert 1\rangle$ are the
eigenvectors of { M}$^{(n)}$ with eigenvalues 0 and 1 respectively.

 The configuration space of a single random variable $n(R_n)$ : $\phi^{(n)}$ is of rank 2 and is spanned
by these vectors $\vert\uparrow\rangle$ and $\vert\downarrow\rangle$ and { M}$^{(n)}$ is an operator on this space. We consider
the configuration space of {\sl all} the variables $\{n(R_n)\}$ : $\Phi = \prod
^\otimes \phi^{(n)}$.
The augmented space formalism \cite{asf} constructs a Hamiltonian :

\begin{eqnarray}  \widetilde{H} &=&\ \ll \varepsilon\gg \sum_{R_n} P_{n}\otimes { I}\otimes { I} + \varepsilon_1 \sum_{R_n}
P_n\otimes { I}\otimes P^n_\downarrow 
 + \varepsilon_2\sum_{R_n}
P_n\otimes{ I}\otimes T^n_{\uparrow\downarrow}\nonumber\\
& &  -t \sum_{R_n}\sum_{R_m} T_{nm}\otimes\ { I}\otimes I
+\hbar\omega \sum_{R_n} I\otimes{ N}_{n}\otimes I  -\lambda \sum_{R_n}  P_n \otimes \left({ T}_n^+
 + { T}_n^-\right)\otimes I
\end{eqnarray}
                                                                                
\noindent where, 
\begin{itemize}
\item $P^n_\downarrow = \vert \downarrow^n\rangle\langle\downarrow^n\vert$ and
$T^n_{\uparrow\downarrow} = \vert\uparrow^n\rangle\langle\downarrow^n\vert +
\vert\downarrow^n\rangle\langle\uparrow^n\vert$ are the projection and transfer
operators in the configuration space $\Phi$
\item  $\epsilon_1 = (1-x)(\varepsilon_A-\varepsilon_B)$ and $\varepsilon_2 = \sqrt{x(1-x)}(\varepsilon_A-\varepsilon_B)$.
\end{itemize}

 This enlarged Hamiltonian is in the full augmented Hilbert space :~
$\widetilde{\bf H}= {\bf H}^{el}\otimes{\bf H}^{ph}\otimes \Phi$. As in
the case of the phonon space, in this {\sl disorder configuration} space, a general vector is a pattern of $\uparrow$ and $\downarrow$-s. The sequence of sites
where we have a $\downarrow$ is called the cardinality sequence and uniquely describes the configuration.
A typical member of a basis in this augmented space is $\vert R_j\otimes\{n_{R_1},n_{R_2},\ldots\}
\otimes \{R_{k_1},R_{k_2},\ldots\}\rangle$.
                                                                                
The augmented space theorem \cite{asf} states that :
                                                                                
\begin{equation}
\ll G(R,E) \gg = \langle R\otimes\{\emptyset\}\otimes\emptyset\vert\
\left(E \tilde{I}-\widetilde{H}\right)^{-1}\ \vert R \otimes\{\emptyset\}\otimes\emptyset\rangle
\end{equation}
                                                                                
\noindent where $\vert\emptyset\rangle\ \epsilon\  \Phi$ is the null cardinality sequence or one where we have $\uparrow$ everywhere. 
Equivalently, exactly as discussed in section 1, we can construct the above formalism in reciprocal  space
by Fourier transforming the electronic part of the basis 
and keeping the rest unchanged.
So we can write the configuration averaged
Green's function in reciprocal space as :

\begin{equation}
\ll G(k,E) \gg = \langle k\otimes\{\emptyset\}\otimes\emptyset\vert\
\left(E \tilde{I}-\widetilde{H}\right)^{-1}\ \vert k \otimes\{\emptyset\}\otimes\emptyset\rangle
\end{equation}

The operation of the terms in the Hamiltonian are mostly identical to (7) except for the second to fourth terms in  equation (8) :
                                                                                
\begin{eqnarray}
\widetilde{H}_2 \vert k\otimes \{n_{R_1},\ldots\}\otimes\{{\bf C}\}\rangle\  & =&\
 \varepsilon_1\ \delta(R_1\in \{{\bf C}\})\ \vert k\otimes \{n_{R_1},\ldots\}\otimes\{{\bf C}\}\rangle\nonumber\\
\widetilde{H}_3 \vert k\otimes \{n_{R_1},\ldots\}\otimes\{{\bf C}\}\rangle\ & =&\
 \varepsilon_2\ \vert k\otimes \{n_{R_1},\ldots\}\otimes\{{\bf C}\pm R_1\}\rangle\nonumber\\
\widetilde{H}_4 \vert k\otimes \{n_{R_1},\ldots\}\otimes\{{\bf C}\}\rangle\ & =&\
-t \sum_\chi \exp\{ik\cdot\chi\} \vert k\otimes \{n_{R_1}-\chi \ldots\}
\otimes \{{\bf C}-\chi\} \rangle\nonumber\\
\end{eqnarray}
                                                                                
\noindent here, $\{{\bf C}\}$ is a general cardinality sequence $\{R_k,R_m,\ldots\}$. 

For obtaining the configuration averaged spectral function we carry out recursion with the starting state $\vert k\otimes\{n_{R_1},\ldots\}\otimes\emptyset\rangle\ =\ \vert 1\rangle$ using the
three term recursion similar to (6) but in the full augmented space.
As before, from the orthogonality of the recursive states, we obtain the coefficients $\tilde{\alpha}(k),\tilde{\beta}(k)$ and
averaged Green function is obtained as a continued fraction :
\vskip 0.3cm                                                                                
\begin{equation}
\ll G(k,E)\gg = \frac{1}{\displaystyle E-\tilde{\alpha}_1(k) -
                \frac{\tilde{\beta}^2_1(k)}{\displaystyle
                 E - \tilde{\alpha}_2(k)- \frac{\tilde{\beta}_2^2(k)}
                  {\displaystyle  {{\ddots}\atop{\displaystyle \frac{\beta_{N-1}^2}{\displaystyle E-\alpha_N-\beta_N^2\ \tilde{T}(k,E)}}}
 }}}
\end{equation}
                                                                                
As before the calculation is carried down to a finite number N of steps and then
the continued fraction is terminated by a function $\tilde{T}(k,E)$ as suggested by Beer and Pettifor.
The configuration averaged spectral function and other
quantities of interest are calculated using
the same formalism as discussed in section 1.

\section{The Ground state in a ordered Holstein model}

 In this section we will discuss the results of our ground-state properties
and compare them with the results of two of the most successful recent
numerical works. 

\begin{figure}
\centering
\resizebox{12cm}{7cm}{\includegraphics{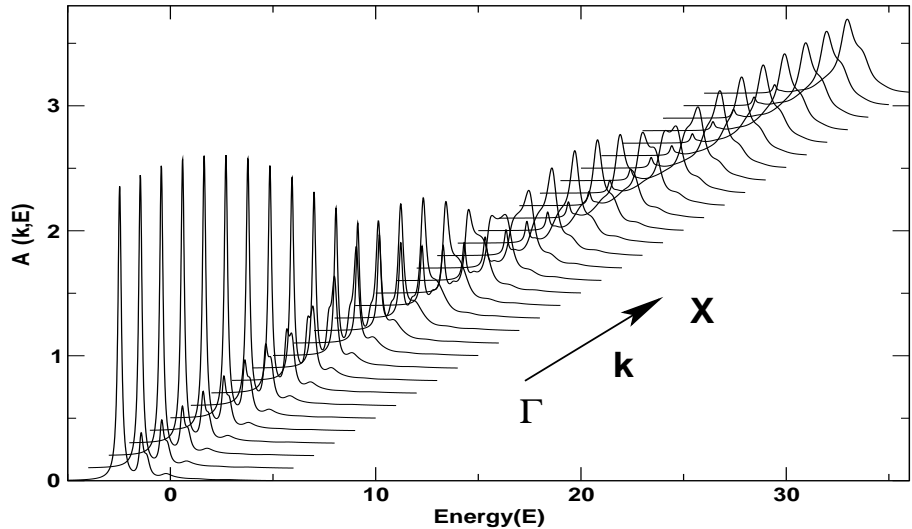}}
\caption{The spectral functions are plotted for different $k$ values. The spectral functions are obtained from a k-space recursion with a 22 shell
nearest neighbour map.}
\label{fig1}
\end{figure}

The figure  \ref{fig1} shows the spectral functions for different $k$ values.
All the spectral functions show a very narrow delta function like peaks at
the lower end of the spectrum.
This is related to the band corresponding to the polaron
ground state. This state has very narrow width, which means this $k$-labeled state has a very long lifetime. The other peaks are related to excited states. These have larger widths.

\begin{figure}
\centering \resizebox{7.5cm}{8cm}{\includegraphics{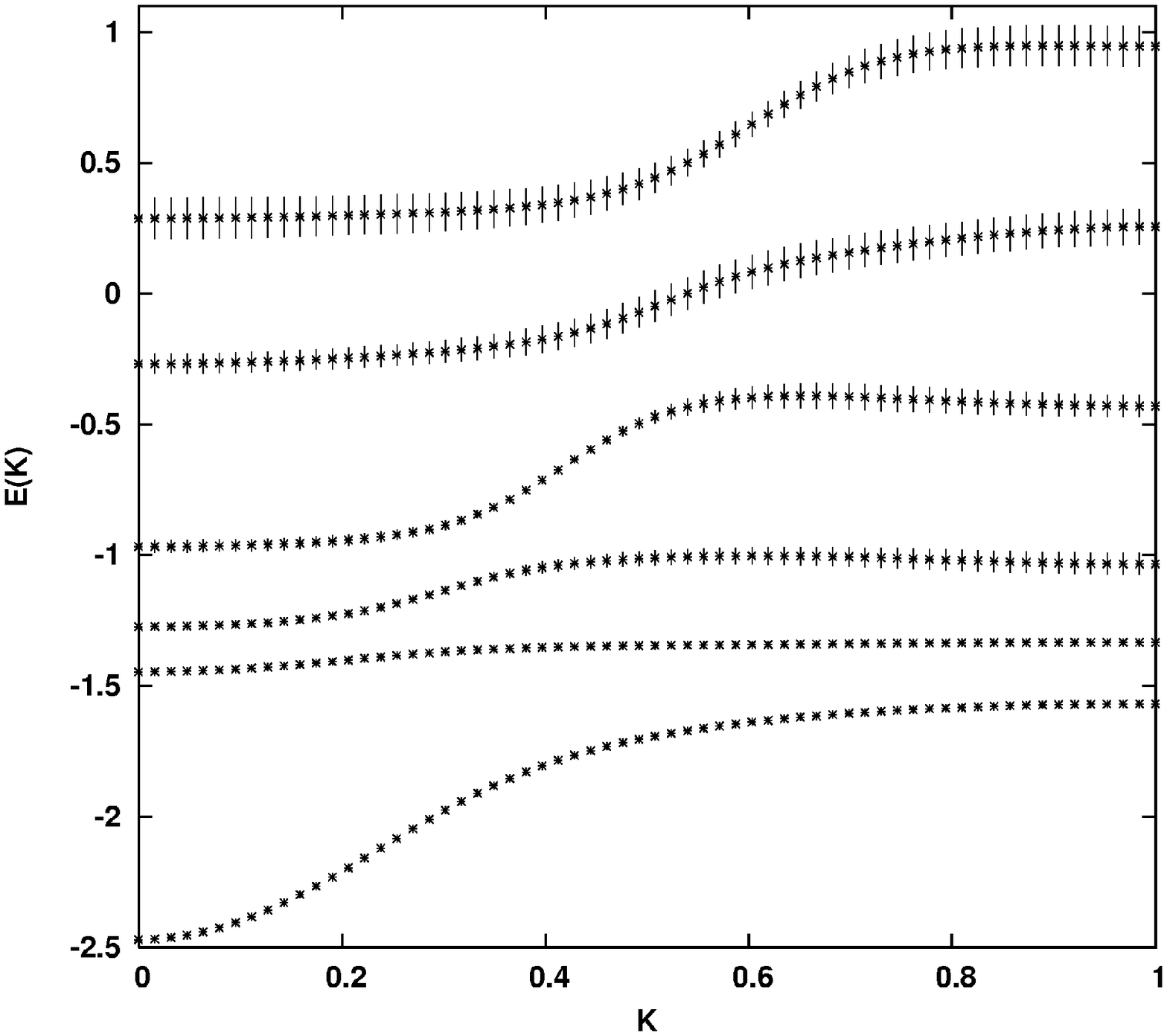}}
\resizebox{7.2cm}{7.2cm}{\includegraphics{grdis.eps}}
\caption{(left) Dispersion curves for ground state and few excited states are
plotted for $\lambda=\omega=1$. The recursion was carried out using a 22 shell map.
(right) Dispersion curves for different values of e-ph coupling.
Here $\lambda/\omega$ = g}
\label{f.1}
\end{figure}

The dispersion curves shown in figure \ref{f.1} (left)  are obtained by fitting Lorentzian's  in the neighbourhood of
 the peaks in the spectral function : the Lorentzian centre gives the energy
$E(k)$ while the width gives the width in the dispersion curves.

    Table-I compares our polaron ground-state energy at k=0 with those 
obtained by VAED \cite{trug1} (in the weak coupling regime)  and DMRG \cite{dmrg1} (in the Strong coupling regime) for two sets of parameters. 
The ground state energies were obtained by two different starting 
{\sl root states} as  mentioned earlier. Our energies 
 match extremely well with  both the earlier VAED and DMRG calculations. 

\begin{table}[b]
\centering
\begin{tabular}{|l|c|c|c|c|c|} \hline
  $\lambda/\omega$ & Present(VAED) & Present(M-VAED) & VAED\cite{trug1} & DMRG\cite{dmrg1} \\   \hline
   1.0 & -2.469684723933 &-2.469684723933  & -2.469684723933 & -2.46968   \\ \hline
    $\sqrt 2$ & -2.998828186866 & -2.998828186866  & -2.998828186867 &-2.99883   \\ \hline
\end{tabular}
\caption{Comparison of Polaron Ground state energy for 
k=0 for two different parameters. M-VAED refers to the phonon enriched starting state described in section 1}
\vskip .4cm
\end{table}
\par

\begin{figure*}
\centering
\vskip .6cm
\resizebox{8cm}{6cm}{\includegraphics{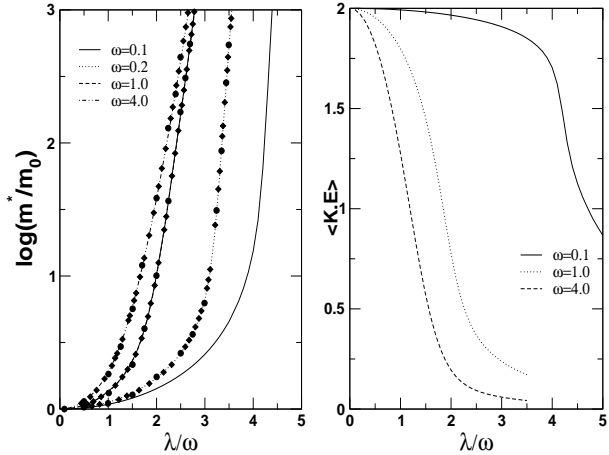}}
\caption{\label{f.4} The left panel shows the logarithm of the ratio 
of the polaronic mass $m^*$ to bare 
electron mass w.r.t. the electron-phonon coupling strength in different Limits. Dots
are the DMRG extracted data points and Squares are VAED extracted data points.
The right panel shows the average kinetic energy for different values of
$\omega$}
\end{figure*}

 We have calculated the effective mass spanning all parametric regimes.
 We find good
agreement with results of the DMRG and the earlier VAED methods. This is shown
in figure \ref{f.4}(left panel).   
To check numerical stability we have done the calculations first with a
17 shell map with a maximum of 34 phonons and another with a 18 shell map
with a maximum of 35 phonons. The results match to within our accuracy window 
 across the parametric regime $0\leq g\leq 5$. 
At the low electron-phonon coupling regime we have a quasi-free electron with a slightly
renormalized mass. As the electron-phonon coupling strength increases the polaron becomes heavier.
 The crossing though always smooth, is  rapid in the adiabatic limit. 
In the same figure(right panel)
 we also plot the averaged kinetic energy, which rapidly
decreases as the polaron becomes heavier .

\begin{figure}[b]
\centering
\vskip 0.7 cm
\resizebox{8cm}{6cm}{\includegraphics{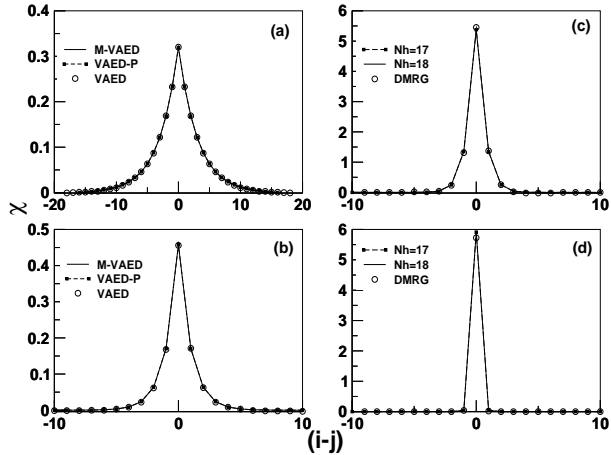}}
\caption{\label{f.5} The Lattice deformation $\chi$ as function of $R_i-R_j$ at k=0 for (a) $\omega$=.1 and$\lambda$=.1, (b)$\omega$=1. and $\lambda$=.5, (c)$\omega$=.1 and$\lambda$=.435, (d)$\omega$=1. and $\lambda$=3. }
\end{figure}

	We have calculated the static correlation 
function $\chi(R_i-R_j)$ between the electron position and 
quanta of lattice vibration which gives
us a measure of the electron-induced lattice deformation and its spatial extent.
We compare our results with VAED and DMRG results. There is excellent
agreement in between our results generated using both of our basis
and VAED results which is shown in figure \ref{f.5}. In figure \ref{f.5}(a)
we show the static correlation function for  
 $\omega$=0.1 (adiabatic) and $\lambda$=0.1(weak coupling), and we can clearly
see spatial extent of the polaron span the system size, therefore a large 
polaron. We do not show the DMRG result in this case as it is less
reliable in this parameter regime.
In figure\ref{f.5}(b) we show the $\chi$ for $\omega$=1.0 and $\lambda$=0.5,
which again matches well with the VAED calculation. 
Figure\ref{f.5}(c) and figure\ref{f.5}(d) 
give us the result for high electron-phonon coupling in 
adiabatic $\omega/t$=0.1
($\lambda$=4.35) and intermediate $\omega/t$=1.0($\lambda$=3.0) limits
respectively. In both the cases
  spatial extent of the polaron has been reduced, resulting in small polaron. Here
we have achieved excellent convergence applying the modified or phonon-enriched
VAED basis (M-VAED), as described in section 1, 
using  N=17 (34 bosons at the root site); N=18 (36 
phonons at root site) and N=19 (37 phonons at the root site) shell maps.
Our energy converges to 11-12 decimal places
for these two calculations. Figure\ref{f.5}(c) shows good agreement with DMRG results.
The value of our extracted DMRG $\chi$(0) is 5.459016. The value of our
 $\chi(0)$ for this case is 5.374, which is very close to 5.4 the lower
 limit of extrapolated VAED data. Figure\ref{f.5}(d)
 also has a good agreement with DMRG results.
The  value of our $\chi$(0) is  5.91 and value of the DMRG extracted data is 5.73.
 For small 
polaron our data are in good agreement with DMRG data. Unlike DMRG results , our results are reflection symmetric (i.e.,$\chi$(l)=$\chi$(-l)) as this basis takes proper care of 
translation symmetry. We conclude that we have achieved highly  reliable  results in all regimes. \\

\begin{figure}
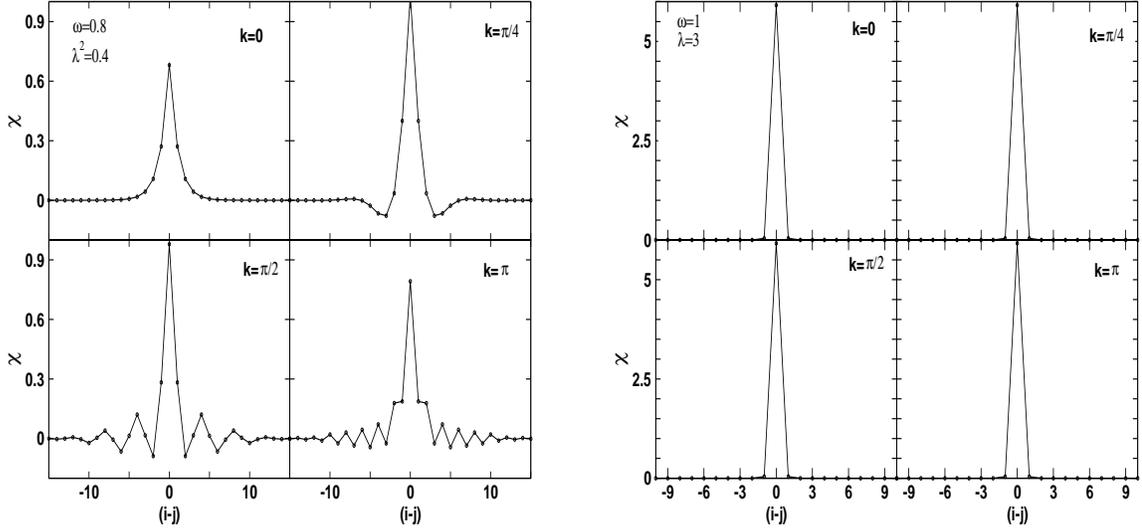

\centering
\vskip 1.3 cm
\resizebox{7cm}{7cm}{\includegraphics{chik1.eps}}\phantom{XXX}
\resizebox{7cm}{7cm}{\includegraphics{chik2.eps}}
\caption{(left) Figure shows lattice deformation $\chi$
for different k values at $\omega=0.8$ and $\lambda^{2}=0.4$. 
 (right) Figure shows lattice deformation $\chi$
for different k values at $\omega=1.0$ and $\lambda=3.0$}
\label{f.c6}
\end{figure}

	We have also calculated $\chi$ at non-zero k at $\omega$=0.8, 
$\lambda^2$=0.4\ (figure \ref{f.c6}, left) and $\omega$=1.0, $\lambda$=3.0(figure
\ref{f.c6},right) to see how the polaron 
transforms from predominantly electronic character at k=$0$ to phononic at
k=$\pi$. In figure \ref{f.c6}(left) at k=0 where group velocity is zero the deformation affects only the sites in its close vicinity , falls off exponentially and is
always positive. At k=$\pi/4$, there is a enhancement in the deformation 
amplitude and it acquires a negative sign in oscillation as well. At k=$\pi/2$ the 
oscillations are enhanced in sign as well as spatial extent. The spatial extent
of the deforming oscillations attains its maximum  at k=$\pi$.
Our calculation matches well with VAED for same set of parameters for zero as 
well as non-zero k-values.

  Figure \ref{f.c6} (right) shows the $\chi$ for same k-values but
for high electron-phonon coupling strength. Here $\chi$ shows hardly any variation with k,
i.e., lattice distortion is predominantly on the electron site, with hardly
any group velocity. Figure \ref{f.1} (right panel)  shows   polaron is almost dispersion less for this parameter
 i.e.,  group velocity(${d\omega}/{dk}$) is almost zero 
and again the average kinetic energy(figure \ref{f.4} right panel) also drops 
 drastically for this value, hinting clearly that a very small 
 polaron has formed.

\begin{figure}
\centering
\resizebox{12cm}{8cm}{\includegraphics{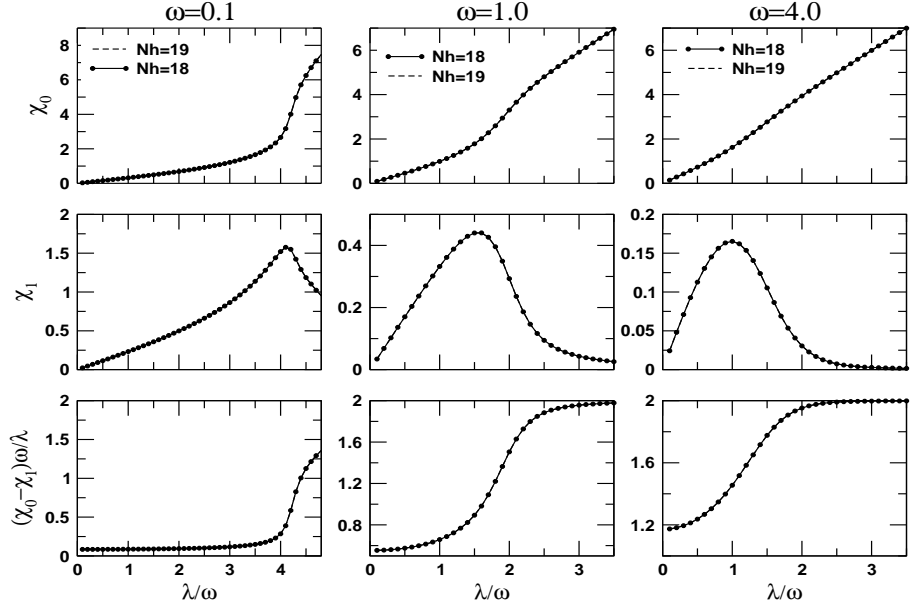}}
\caption{\label{f.6}Figures in the upper row shows the $\chi$(0) at three different $\omega$. Middle row is for $\chi$(1). Lower row is for ($\chi$(0)-$\chi$(1))/g }
\end{figure}

	We have also calculated the correlation functions $\chi$(0), $\chi$(1),
 ($\chi$(0)-$\chi$(1))/g which is shown in figure \ref{f.6}. Where g is $\lambda$/$\omega$. $\chi$(0) has a non-linear behavior
in the intermediate electron-phonon coupling regime. It is linear w.r.t g both in the low and high electron-phonon
coupling regime and this trend is there in all limits(i.e., for different $\omega$). But slope of $\chi$(0) 
in the low electron-phonon coupling regime depends on $\omega$ but its slope in high electron-phonon coupling regime
becomes independent of $\omega$ and its value approaches 2. The change in
 slope of $\chi$(0) as we go from low  to high  g, is more prominent for lower values of $\omega$. $\chi$(1) increases linearly to start with(i.e., low g regime), then its rate of increment decreases
and then it starts decreasing, signaling the arrival of high electron-phonon coupling 
regime. The decrement in  $\chi$(1) w.r.t to g and the corresponding change in 
$\chi$(0) implies that, from this g onward the lattice distortion starts
 getting confined to the electron site.
The correlation function ($\chi$(0)-$\chi$(1))/g substantiates these facts. It 
always saturates to a value 2 for high g regime. For $\omega$=0.1 we have not 
gone to that magnitude of g where 
($\chi$(0)-$\chi$(1))/g would saturate to 2, but the trend is quite clear. For 
lower values of $\omega$ the cross-over is rather sharp, which is also 
reflected in the Effective mass and average kinetic energy calculation (figure \ref{f.4}). 

\section{The First Excited-State in an ordered Holstein model }
   
In this section we discuss our calculation for the first excited 
state of the electron-phonon system. The first excited state consists of the ground state
polaron and an unbounded extra phonon excitation\cite{trug1}. Here the
energy difference between the ground state and the first excited state should
be equal to $\omega$ for a infinite chain(i.e., thermodynamic limit) and the
mean phonon number difference should be equal to one(i.e., $\Delta N^{ph}$=1.0). But for
the excited state there are two distinct regimes, one below the critical electron-phonon
coupling $\lambda_c$ where the excited phonon tries to be at a infinite distance away
 from the ground state polaron(but the finite size of the system hinders it)and above it
where the excited phonon is absorbed by the ground state polaron forming a
bound state. The calculations here are done for $\omega$=0.5. VAED\cite{trug1} 
calculations showed that a phase-transition  occurs in the first excited state 
at $\lambda_c$=0.95(i.e., g=1.9) for $\omega$=0.5. Our calculations  shows 
a phase-transition precisely at $\lambda$=0.95. Our calculated binding energy ( for the first excited state) and  correlation  functions  clearly answers to the issue, whether 
the extra phonon excitation  forms a bound state with the ground state polaron 
after a certain point(i,e., $\lambda_c$)  or prefers to remain infinitely 
 separated(of course limited by the finite size  of the system) for all values
 of $\lambda$.
 \par

\begin{figure}
\centering
\vskip 1cm
\resizebox{8.8cm}{7.0cm}{\includegraphics{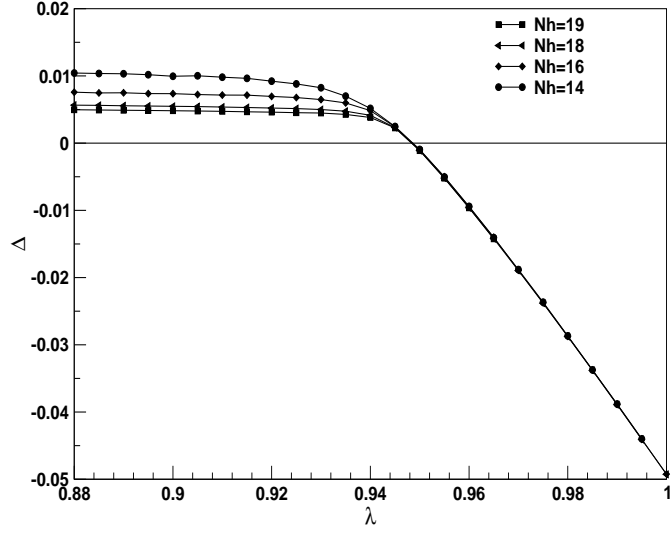}}
\caption{Figures shows the binding energy as a function of
$\lambda$ for different system sizes(i.e., using different shell maps).}
\label{fig4}
\end{figure}

  The binding energy $\Delta=\Delta E$ - $\omega$(where $\Delta E=E_1-E_0$,
$E_1$ and $E_0$ are the first-excited-state and 
ground-state energy respectively) as a function of 
electron-phonon coupling $\lambda$ is shown in figure\ref{fig4} for various
system sizes. Below  $\lambda_{c}$, $\Delta$ varies with system size but is 
greater than zero. This variation with system size is due to the fact, that the
excited phonon want to be infinitely away from the the ground state polaron,
but the system size limits its ability. As the system size increases it
slowly approaches the thermodynamic limit, i.e., $\Delta$=0. For $\lambda>\lambda_{c}$, where the absorption of the excited phonon by the ground-state-polaron
has resulted in formation of a bound state, $\Delta$ has clearly converged at
Nh=14 and is negative. 

\begin{figure*}
\centering
\vskip 1cm
\resizebox{12cm}{7cm}{\includegraphics{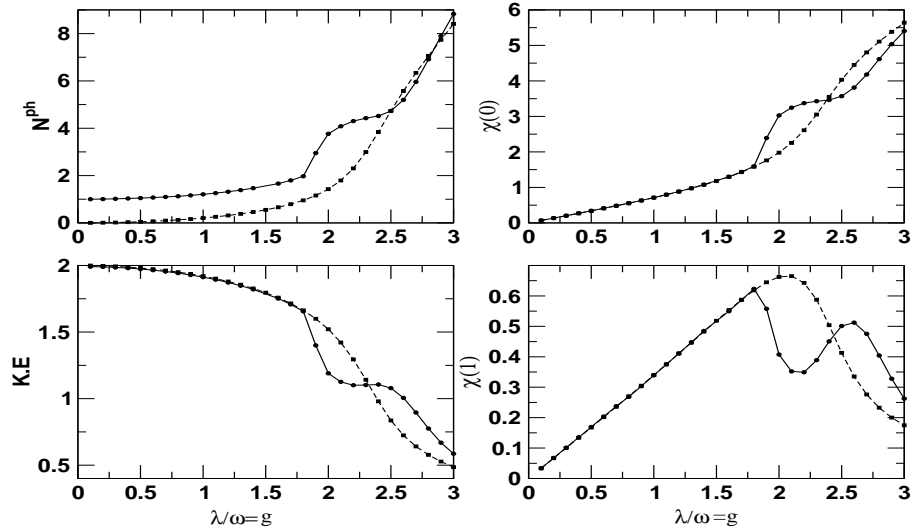}}
\caption{\label{f.7}Figures show the comparison between the first excited state and ground 
state Correlation Functions at $\omega$=0.5. Phase transition in the first 
excited state occurs at $\gamma$=0.95 i.e., g=1.9. Solid line with dots
are results for the  first excited state and dashed line with squares are results
for the  ground state.}
\end{figure*}

   Figure\ref{f.7} compares the different correlation functions of the first excited state and
 the ground state for $\omega$=0.5. There is steady difference in between  $N^{ph}$ of
the first excited and the ground state($\Delta N^{ph}\simeq$1.0 ) below g=1.9, and then
 starts diverging. They again appears to converge at higher g though our excited state
  are not very accurate for higher g. Average kinetic energy, $\chi$(0), $\chi$(1) are
same for the ground state and the excited state below $\lambda_{c}$ which very clearly
indicates that below $\lambda_{c}$ the excited phonon and the ground state polaron are
unbounded. The average kinetic energy, $\chi$(0), $\chi$(1) of the first excited state
below $\lambda_{c}$ are entirely that of the ground state polaron and it is transparent
to the presence of the excited phonon\cite{trug1}. At $\lambda_{c}$ the root site 
disassociates itself from the rest of the lattice\cite{trug1}, the signature of which
can be found in the sudden rise in $\chi$(0) and corresponding fall in  $\chi$(1) .
 The bound state formed by the absorption of the excited phonon by the ground state 
polaron is a excited polaron. It slowly stabilizes with increasing $\lambda$ and exhibits
the behavior of a de-excited polaron.
\par

\begin{figure}
\centering
\vskip 1cm
\resizebox{12cm}{7cm}{\includegraphics{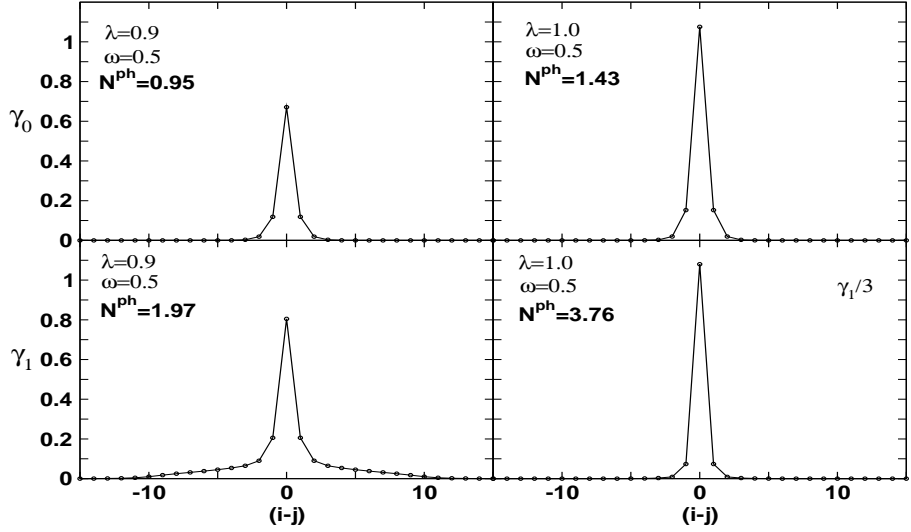}}
\caption{Figures shows the phonon number $\gamma$ as a function of the distance
from the electron position $R_i-R_j$.  }
\label{fig12}
\end{figure}

   In figure \ref{fig12} we have calculated the distribution of the number of excited phonons in
the vicinity of the electron $\gamma(R_i-R_j)$. We show  $\gamma(R_i-R_j)$ both for ground state
($\gamma_0$) and and excited state($\gamma_1$) just above($\lambda$=1.0) and below
($\lambda$=0.9) the transition point. Below the transition point the peaks of $\gamma_{0}$
and $\gamma_{1}$ are almost same , but $\gamma_{1}$ has a longer tail suggesting that tail
represent the extra excited phonon which extends throughout the system in its attempt
to remain unbounded from the ground state polaron. Our calculated $N_{ph}$ are same as
VAED\cite{trug1} for all the cases. The difference in between the ground state $N_{ph}$
and first excited state $N_{ph}$ should be one for a infinite system, but our difference
is about 1.02 and this can be attributed to finite size effect\cite{trug1}. The situation
above $\lambda_c$ is different. Here the peak value of $\gamma_1$ is almost three times the
peak value of $\gamma_0$ and secondly here  $\gamma_1$ decays fast unlike that below
$\lambda_c$. Here the difference in phonon number of the ground state and the excited
state is 2.33 again exactly same as VAED\cite{trug1}. Thus here the excited bound polaron
has many extra phonon excitations compared to the ground state polaron. Since the extra
phonon excitation are almost confined to the root site of the electron, it throws some
light on the fact that at $\lambda_c$ the root site gets detached from the rest
of the lattice\cite{trug1}.

\vskip 0.50cm

\section{Results for a disordered Holstein model}

\begin{figure}[h!]
\centering
\resizebox{10.0cm}{6.0cm}{\includegraphics{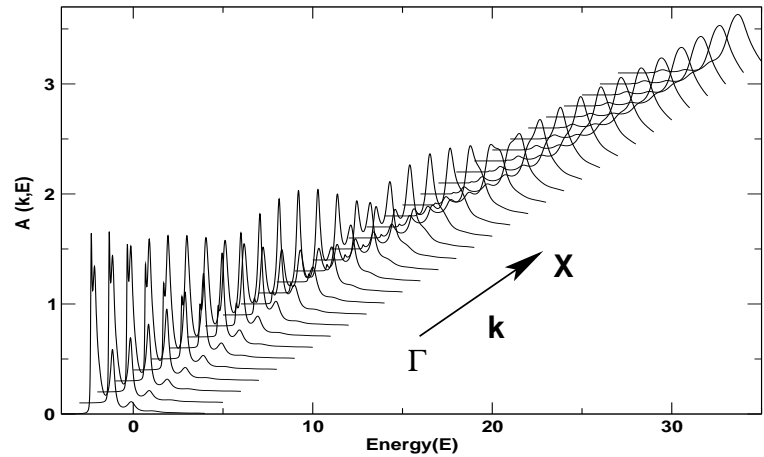}}
\caption{The spectral functions for the disordered Holstein model}
\label{figsp}
\end{figure}

\begin{figure}[h!]
\centering
\vskip 1.0cm
\resizebox{15.0cm}{10.0cm}{\includegraphics{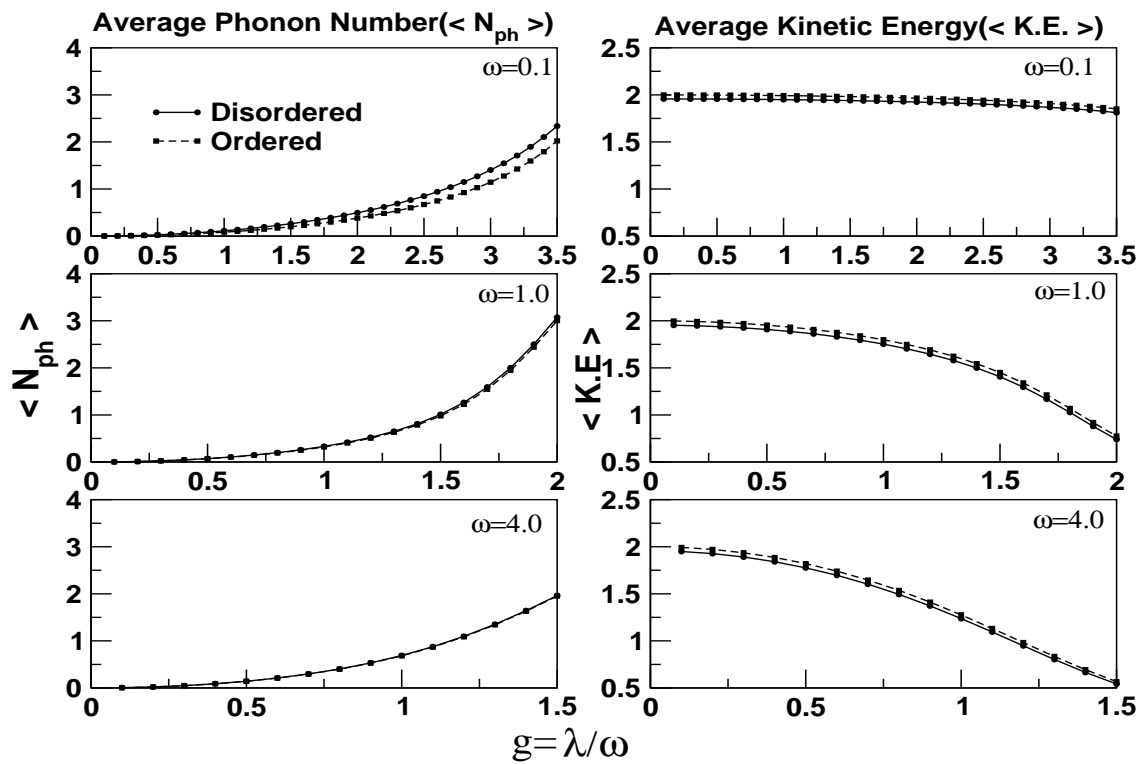}}
\caption{Comparison of the average phonon number and kinetic energy for the disordered and ordered Holstein model at different oscillator frequencies}
\label{fig_d1}
\end{figure}

\begin{figure}[h!]
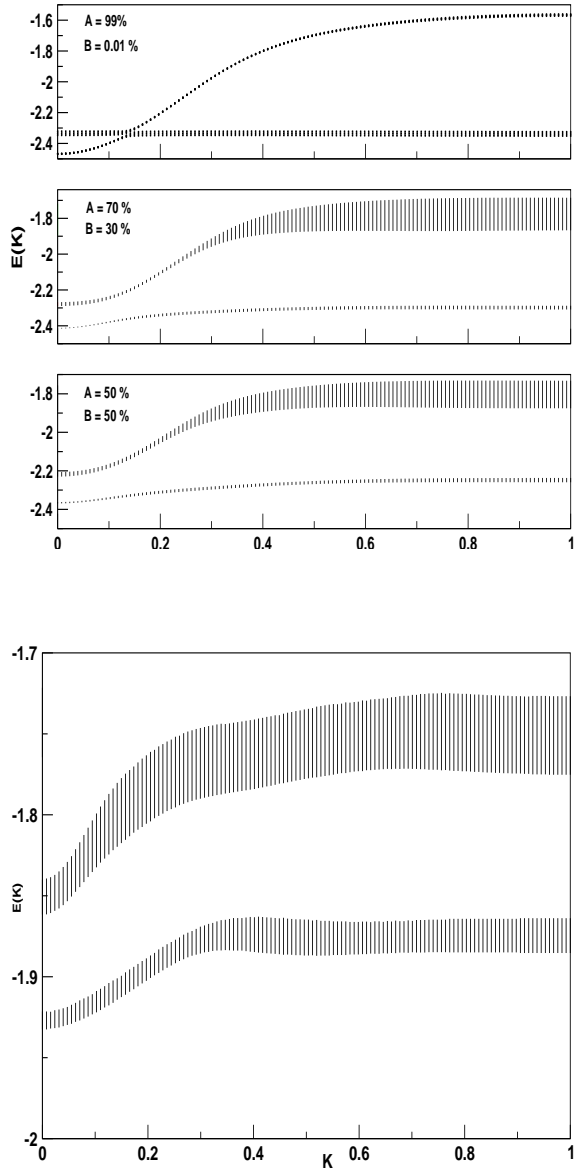

\centering
\vskip .5cm
\resizebox{7.5cm}{7.5cm}{\includegraphics{dis.eps}} 
\vskip 1cm
\resizebox{7.5cm}{7cm}{\includegraphics{lowg.eps}}
\caption{(Top) The dispersion curves for the disordered Holstein model, for g=1,
$\epsilon_A$=0, $\epsilon_B$=0.5, for three different compositions.
 (Bottom) The dispersion curves for the disordered Holstein model, for g=0.01,
$\epsilon_A$=0, $\epsilon_B$=0.5, for the 50-50 alloy.}
\label{figsp2}
\end{figure}

\begin{figure}[h!]
\centering
\resizebox{15.0cm}{10.0cm}{\includegraphics{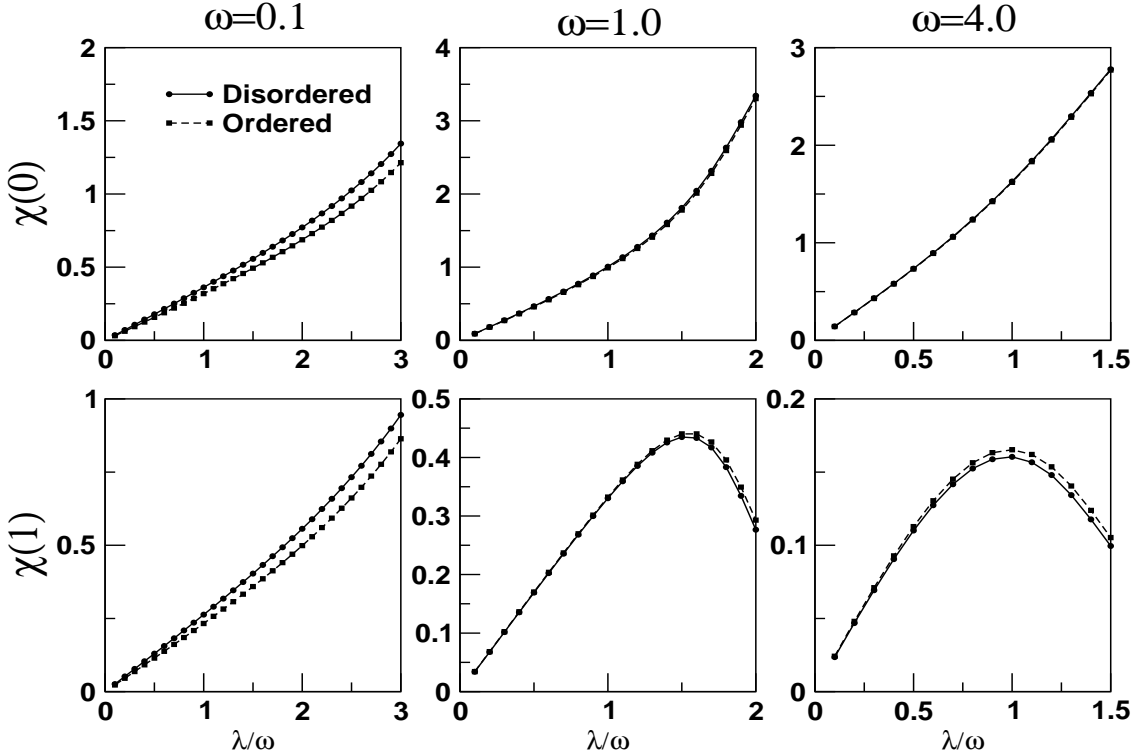}}
\caption{Comparison of $\chi$(0) and $\chi$(1) for the disordered and ordered Holstein model at different oscillator frequencies}
\label{fig_d2}
\end{figure}
\newpage

In figure \ref{figsp} we show the spectral function from the $\Gamma$ to $X$ point in  Brillouin zone  for the
50-50 alloy. The spectral functions show the extra disorder induced widths as well as contributions coming from both the components. This is much better seen in the dispersion curves shown in figure \ref{figsp2}(top).
The dispersion curves for both the disordered alloys show the two branches
arising out of the two components and the large disorder induced widths in the
upper band. In the low concentration impurity regime the two bands almost cross each
other. For higher concentrations the bands are well separated. As in the ordered
case, the lower band has very little width. 
 In comparison
with the ordered case the upper band has large width particularly near the Brillouin zone edge. This extra width arises from quenched disorder scattering. These dispersion curves are for g=1. Figure \ref{figsp2} (bottom) shows the

dispersion curves for g=0.01, i.e. low electron-phonon coupling. 
Here the disorder effect dominates and {\it both} the branches show disorder induced large
smearing. For low electron-phonon coupling in a disordered alloy, the polaron, when formed, will have disorder scattering induced finite life-time. 
We conclude that disorder scattering effects are small 
on heavy polarons, while they tend to give larger 
lifetime effects to light polarons.

            We have also calculated different ground state correlations 
functions for the disordered Holstein model and have compared it with its ordered
counterparts to get a better insight of the disorder effect.
 Figure \ref{fig_d1} compares the average phonon number and the average 
kinetic energy for the ordered and disordered case for three oscillator frequencies.
Site disorder tends  enhances the average phonon number with increasing 
electron-phonon coupling in the adiabatic regime where as, in the intermediate
 and  anti-adiabatic regime disorder has hardly any effect. Disorder lowers 
the average kinetic energy in all the regimes. Figure \ref{fig_d2} 
compares the $\chi$(0) and $\chi$(1) and similar trend is observed in this`
set of correlation functions too. We conclude that the disordered effect is 
more prominent in the adiabatic regime and with increase in the oscillator 
frequency, the phononic disorder becomes dominant.  
\newpage
We would like to acknowledge effective discussions with Prof. A. N. Das and
Dr. J. Chatterjee of SINP, Kolkata and Dr. P. A. Sreeram of SNBNCBS.



\begin{thebibliography}{99}
\bibitem{her} A function of a complex variable f(z) is called herglotz if : Im f(z) = -sgn[Im(z)] and f(z)$\rightarrow 0$ as z$\rightarrow \pm\infty$ along the real z line.
\end{thebibliography}
\end{document}